\documentclass[prb,preprint,preprintnumbers,amsmath,amssymb]{revtex4-1}

\usepackage{graphicx}
\usepackage{dcolumn}
\usepackage{bm}
\usepackage{bibtopic}

\begin{document}


\def\Ef{$E_{\rm F}$}
\def\Ed{$E_{\rm D}$}
\def\Eg{$E_{\rm g}$}
\def\Efmath{E_{\rm F}}
\def\Edmath{E_{\rm D}}
\def\Egmath{E_{\rm g}}
\def\Tc{$T_{\rm C}$}
\def\kpara{{\bf k}$_\parallel$}
\def\kparamath{{\bf k}_\parallel}
\def\minuskpara{$-{\bf k}_\parallel$}
\def\kparazero{{\bf k}$_{\parallel,0}$}
\def\minuskparazero{$-{\bf k}_\parallel,0$}
\def\kperp{{\bf k}$_\perp$}
\def\invA{\AA$^{-1}$}
\def\Kbar{$\overline{\rm K}$}
\def\Gbar{$\overline{\Gamma}$}
\def\Mbar{$\overline{\rm M}$}
\def\GbarMbar{$\overline{\Gamma}$-$\overline{\rm M}$}
\def\GbarKbar{$\overline{\Gamma}$-$\overline{\rm K}$}
\def\DeltaEf{$\Delta E_{\rm F}$}
\def\DeltaEfmath{\Delta E_{\rm F}}
\def\BiTe{Bi$_2$Te$_3$}
\def\BiSe{Bi$_2$Se$_3$}
\def\BiSnTe{(Bi0.67\%Sn)$_2$Te$_3$}

\title{ High spin polarization and circular dichroism of topological
                 surface states on \BiTe\   }
  \author{ M. R. Scholz,$^1$, J. S\'anchez-Barriga$^1$, D. Marchenko,$^1$ A. Varykhalov,$^1$    
           A. Volykhov,$^2$ L. V. Yashina,$^2$ and O. Rader$^1$ }
 \address{$^1$Helmholtz-Zentrum Berlin f\"ur Materialien und Energie, 
Elektronenspeicherring BESSY II, Albert-Einstein-Str. 15, D-12489 Berlin, Germany}
 \address{$^2$Department of Chemistry, Moscow State University, 
          Leninskie Gory, 1/3, 119992 Moscow, Russia}
  
\begin{abstract}

Topological insulators have been successfully identified by spin-resolved 
photoemission but the spin polarization remained low ($\sim20$\%).
We show   for   \BiTe\ that
the   in-gap surface state  is much closer to full spin polarization
with measured values reaching $80$\%\ at the Fermi level. 
When hybridizing with the bulk it remains
highly spin polarized which may explain
 recent unusual quantum interference results on \BiSe.
The topological surface state shows a large circular dichroism in the 
photoelectron angle distribution with an asymmetry of $\sim20$\%\
the sign of which corresponds to that of the measured spin. 

\end{abstract}


\maketitle

The   topological insulator 
\cite{KanePRL05,BernevigPRL06,Koenig07,FuPRL07,MoorePRB07,Roy06,Murakami,HasanBiSbSpin,WellsPRL09,HsiehNature09}
was proposed in 2005 when 
  its two-dimensional version, the   quantum spin Hall 
insulator, was  investigated taking the example of graphene \cite{KanePRL05}:
Provided a sizeable spin-orbit interaction existed in that material, 
a corresponding electronic band gap opens at the otherwise gapless 
Dirac crossing point. 
If the sample geometry is chosen to be that of a one-dimensional graphene ribbon,
then  inside of the gap only the two-dimensional bulk states are forbidden but
at the one-dimensional boundaries of the ribbon, 
 edge states will cross this  gap \cite{KanePRL05}.
Due to time-reversal symmetry, these states form Kramers pairs of
opposite spin and   opposite propagation directions.
 These     chiral Dirac fermion states are  
100\%\ spin polarized, and the carriers cannot backscatter
without violating spin conservation \cite{footnoteA}.
Because the spin-orbit interaction was not large enough in graphene \cite{KanePRL05},
alternative systems were sought and identified in theory \cite{BernevigPRL06}.
The quantum spin Hall phase was eventually realized in HgTe/Hg$_{1-x}$Cd$_x$Te 
quantum wells
where one-dimensional and chiral   edge states could be detected by their
transport properties \cite{Koenig07}. 

Subsequently,   three-dimensional  topological insulators have been predicted 
where the spin-orbit gap 
is the gap of the three-dimensional
bulk states and     topological protection of the 
two-dimensional surface state is again guaranteed by time-reversal symmetry
\cite{FuPRL07,MoorePRB07,Roy06,Murakami}. 
This prediction has particularly well been received in the surface science
community because, on the
one hand, angle-resolved photoelectron spectroscopy permits the direct measurement of 
two-dimensional surface states   and, on the other hand, 
the concept shares important properties with the spin-orbit-driven Rashba effect
which is currently under intense investigation in semiconductor heterostructures
 \cite{Nitta} 
as well as at metal surfaces   \cite{LaShellPRL96,Hirahara,Ast}
and quantum wells  \cite{VarykhalovQWS}.
In fact, the chiral spin arrangement where the spin is 
  tangential to the constant-energy surfaces is similar
in the Rashba effect while the topological protection and  single-spin character
of the surface state are characteristic of  the topological insulator
[Fig. 1(a,b)].

A topological surface state is present if it crosses the Fermi energy
 (\Ef) an odd number of times between two time-reversal invariant points 
of the surface Brillouin zone \cite{FuPRL07}, and for a proper counting,
spin resolution was shown to be indispensable 
\cite{HasanBiSbSpin,WellsPRL09,HsiehNature09}. 
Among the systems predicted to be topological insulators 
are \BiSe\ and \BiTe\ \cite{FuPRL07,ZhangNatPhys09} 
with trigonal crystal structure  [Fig. 1(d)] and 
 bandgaps large enough to enable applications in spin-dependent transport at
room temperature. These materials are electronically simple 
with just one Dirac cone predicted \cite{FuPRL07,ZhangNatPhys09}
and measured  at \Gbar\ \cite{HsiehNature09,XiaNatPhys09,ChenScience09}. 
Spin-resolved photoemission confirms
that the    surface state of \BiTe\ is spin polarized and  the spin polarization 
reverses indeed between two-dimensional electron wave vectors 
$+$\kpara\ and $-$\kpara\  \cite{HsiehNature09}.  

When backscattering of carriers occurs at the surface, it involves a transition 
from $+$\kpara\ to $-$\kpara\ and thus  would violate spin conservation  
if states at $+$\kpara\ and $-$\kpara\ are of opposite spin \cite{KanePRL05}.
There is naturally high  interest  in proving and using this avoided 
backscattering
in electron transport but this has met with difficulties
\cite{TZhangPRL09,CheckelskyPRL09,ButchArxiv10,RoushanNature09,AnalytisPRB10}.
On a local scale, scanning tunneling microscopy (STM) shows that electrons
occupying the topological surface states are not 
backscattered by nonmagnetic surface impurities \cite{TZhangPRL09}. 
On the other hand, these  electrons do backscatter from surface steps 
\cite{TZhangPRL09,RoushanNature09}.
Recently, Landau levels of the topological surface state of \BiSe\ samples
with a mobility of $\sim 0.1$ m$^2$/(Vs) were observed by STM
but only upwards from the Dirac point \cite{PChengarXiv10}.
On a macroscopic scale, undoped high mobility [$>2$ m$^2$/(Vs)] \BiSe\ samples
showed a surface mobility less than the bulk one \cite{ButchArxiv10}.
\BiSe\ thin film samples showed only bulk transport
 down to 50 nm thickness despite the simultaneous presence of the
Dirac cone in photoemission \cite{AnalytisPRB10}.
Doping of \BiSe\ by Ca can overcome its $n$-type self-doping and move \Ef\ 
into the bulk gap \cite{CheckelskyPRL09}.
In quantum interference experiments, a spin-based magnetofingerprint, i. e., conductance
versus magnetic field, was measured which was found inconsistent with surface states
but consistent with bulk states provided that those share a similar spin arrangement 
with the topological surface state \cite{CheckelskyPRL09}.
On compositionally modulated bulk-insulating \BiTe\ samples, 
Shubnikov-de Haas oscillations have most recently 
been measured giving surface
mobilities of $\sim1.0$ m$^2$/(Vs) \cite{QuScience10}.
 
 At the present stage,  several questions    remain  open. 
The spin polarization of the topological surface states of \BiTe\
has been measured by photoemission at \Ef\ \cite{HsiehNature09}. It shows a
  peak and reverses between $+$\kpara\ and $-$\kpara, as expected,
but the spin polarization reaches  only $\sim20$\%\ which is unlikely 
caused by overlapping $+$ and $-100$\%\ polarized peaks as has been suggested 
\cite{HsiehNature09}. 
We investigate at first whether the spin polarization is intrinsically limited
or whether it is only subject to extrinsic effects that can be technically
  overcome. 

Single crystals of \BiTe\ and \BiSe\ were grown from melt by the Bridgman method.
The entire 10 cm long  crystals could be cleaved along their length, indicating 
the high quality and absence of twinning deformations.
The growth time including cooling was about 2 weeks for a 40--50 g crystal.
Using an adhesive tape, surfaces have been prepared in 
ultrahigh vacuum by {\it in situ} cleavage 
along the trigonal axis   which produced 
(00$\cdot$1)   surfaces in hexagonal coordinates [(111) in 
rhombohedral coordinates]. 
$E({\bf k}_\parallel)$ dispersion relations and constant energy surfaces 
have been measured 
with a Scienta SES100 electron energy analyzer at the UE112-PGM2a beamline 
of BESSY II
with $s$, $p$, $\sigma+$, and $\sigma-$   polarized undulator radiation. 
The incidence angle $\beta$ was 45$^\circ$. 
At the PGM1 branch, spin- and angle-resolved photoemission 
has been measured with linearly polarized light.
The spin-resolved experiment measures the spin component in 
the surface plane of the sample 
and perpendicular to \kpara\ and averages over a sample area of the order of 
500$\mu$m$\times$ 500$\mu$m and over a \kpara\ range of 
$\pm0.025$ \invA. 
We found that cutting the samples perpendicular to the cleavage plane should 
be done without applying pressure so that the damage is limited to the periphery 
of the sample. This can be seen by changes in the low-energy electron
diffraction (LEED) and a position dependence in angle-resolved photoemission. 
Control of this together with the successful in situ cleavage was 
found to be the main factor in obtaining high spin polarization,
and we believe that our lower values for \BiSe\ are also due to 
imperfect samples. 
 
Figure 1c indicates with an arrow the van-der-Waals-type (00$\cdot$1)  cleavage plane, 
which we characterized by LEED   at 80 eV [Fig. 1(e)]. 
Figure 1e shows a characterization of \BiTe\ by angle-resolved photoemission
where Fig. 1(f) (top) shows  the topological surface state (TSS), the Dirac point
(\Ed),
and the valence band maximum (VBM). 
Different from the simple topological insulator model \cite{KanePRL05}, 
the Dirac point in 
\BiTe\ lies in the valence band which in principle would enable
 the topological surface state to couple to bulk states. 
The position of the conduction band minimum (CBM) (which does not appear here 
due to the \kperp\ respectively photon-energy dependence of the 
photoemission transition) indicates that the probed
sample region is  $n$-doped in agreement with our Hall-effect measurements. 
Along \Gbar-\Mbar,  
the linear dispersion of the surface state changes within 100 meV of \Ef\ to
a dispersion with smaller group velocity. This is related to the warping of the 
constant energy surface away from the circular shape at 100 meV binding 
energy seen in Fig. 1(g). It confirms previous measurements for \BiSnTe\ 
and \BiTe\  \cite{ChenScience09} 
and investigations in theory  \cite{FuPRL09}. 
The warping is predicted to lift the prohibition of backscattering  \cite{FuPRL09}. 
Within 100 meV from \Ed, the constant energy surface
is, however, isotropic. 
The data in Fig. 1(f) (bottom) was measured under the same conditions as in Fig. 1(f) (top)
but with  s-polarized light which confirms the
even symmetry of the topological surface state and of the bulk state below
\Ed.  

Figure 2a shows spin- and angle-resolved photoemission data for 
\BiTe\ at 50 eV photon energy. 
The data was measured at $\kparamath=0.11$ \invA\ 
and for \BiSe\ in Fig. 2b at 0.09 \invA\ approximately following \Gbar-\Kbar. 
We show the  photoemission intensity for spin up ($I^\uparrow$) 
and spin down ($I^\downarrow$) along with the detected spin asymmetry
[$(I^\uparrow-I^\downarrow)/(I^\uparrow+I^\downarrow)$]. 
The spin-resolved count rate from \BiTe\ and \BiSe\ is rather low 
even as compared to
the background counts so that the spin asymmetry  is reduced accordingly.
Therefore, we determine above \Ef\ the background intensity $I_B$ (horizontal line
in Fig. 2) and plot the intrinsic spin polarization 
[$(I^\uparrow-I^\downarrow)/(I^\uparrow+I^\downarrow-2I_B)$]. 
(Note that some extremely large error bars in the spin polarization result 
merely from formally negative intensities after the background subtraction.)
We see that   the topological surface state of \BiTe\ is  close to 100\%\
spin polarized [$(82\pm10)$\%\ spin polarization for 
binding energies 0.02 to 0.18 eV].
Many cleavages, however, resulted in lower spin polarization around 50\%,  
and for \BiSe\ we did not obtain cleavages yielding more than 25\%\ for
the topological surface state, most likely due to sample imperfections. 
(Note that the sign of the spin polarization in Fig. 2b is reversed because
also the emission angle is reversed with respect to Fig. 2a.)
The fact that a spin polarization not too far from  100\%\ can be measured means that
this quantity is not principally limited to values around 20\%\ measured previously,
and it confirms the previous interpretation as topological surface state
\cite{HsiehNature09}. 


In Fig. 2 we have marked besides the TSS also the bulk state (BS)
which is connected to it through the Dirac point [see Fig. 1(f)].
For \BiTe, the band structure calculation \cite{XiaNatPhys09} 
assigns BS unambiguously to bulk. 
For \BiSe, BS means rather bulk and/or surface because the calculation 
\cite{XiaNatPhys09} shows a perfect lower half of the Dirac cone 
of the topological surface state while in experiment it is filled
with  bulk states \cite{ZhangNatPhys09}. 
For both systems the states BS (as well as BS$_2$) are 
highly spin polarized with a polarization of $-22$\%\ for \BiTe\ 
(36\%\ for BS of \BiSe\ in Fig. 2b) and a reversal of the spin at \Ed\ 
as expected for the topological insulator. 
TSS and BS can and must hybridize as they share the same symmetry
[see Fig. 1(f)]. 
As a bulk state is principally unpolarized due to bulk inversion
symmetry within the quintuple layer [Fig. 1(d)], this means for \BiTe\ that
bulk states are  below \Ed\ hybridized with the topological surface state 
and in this way can acquire a spin polarization \cite{Fussnote}.
For \BiSe\ possibly the same 
happens. Therefore, the postulation of a strong hybridization of topological 
surface states with bulk states under transfer of the locking of spin and 
momentum direction to explain the magnetic fingerprint data of \BiSe\ is 
not unrealistic \cite{CheckelskyPRL09} but at least transport below \Ed\ 
must be characterized by degenerate TSS and bulk states. 
Note that an unusual coupling of topological surface states with bulk states 
was also reported for Bi10\%Sb \cite{TaskinPRB09}. 

We also applied circular dichroism in the angle distribution (CDAD) of
photoelectrons to \BiTe. In this case, a spin polarimeter is not required 
and, instead, photoemission spectra are recorded subsequently for $\sigma+$ 
and $\sigma-$ light.
It is important to note that the circular dichroism does not directly
couple to the electron spin. This distinguishes this method from spin-resolved 
photoemission. In an atomic description, the spin effect enters via 
the spin-orbit coupling since 
the light interacts with the orbital angular momentum.
Generally, a circular dichroism effect occurs if the system and the
photoemission geometry [Fig. 1(c)] with $\sigma+$ and $\sigma-$ light
are inequivalent. In the present case, the symmetry of the system is
lowered by the spin-orbit interaction. 
The valence-band photoemission is described by a matrix
element with initial- and final-state wave functions. 
The matrix element depends on how the double-group representation
of the dipole operator which includes the direction of circular 
polarization relates to those of the initial and final state which
include the spin. In ferromagnetic transition metals, 
spin-dependent circular dichroism effects of up to 
5\%\ dichroism asymmetry [$(I_{\sigma+}-I_{\sigma-})/(I_{\sigma+}+I_{\sigma-})]$ 
have been measured 
\cite{Bansmann,KuchSchneider,footnote3}.

Figure 3a shows that for $\sigma+$ light  
TSS  is more intense for positive than for negative emission angles, while for
$\sigma-$ light it is the opposite. The dichroism effect amounts to 
$20$\%\ (e. g., at 0.1 eV binding energy). 
When we follow the TSS through the Dirac point into BS, we see that the
dichroism asymmetry reverses exactly as the spin does in Fig. 1 and 2. This is an indication
that the present measurement may probe the angular momentum to which 
the spin is coupled through strong spin-orbit coupling.  
Note that  in \BiTe, the spin-orbit interaction is much larger  than for the 
ferromagnetic transition metals due to higher atomic numbers. 

As in the present work we combine the methods of spin- and angle-resolved
photoemission and of circular dichroism in the angle dependence of photoemission,
we can confirm that the observed large circular dichroism of the topological 
surface state and the connected bulk state BS does correspond to a large spin polarization
and its wave-vector dependence. 
We cannot, at present, prove that the spin {\it via} the angular momentum is 
the reason of the strong circular dichroism. 
Also band-like states of light elements ($Z=6$) have shown large circular dichroism
effects in the angle distribution \cite{SchoenhenseEPL92}.
In this situation, only a photoemission calculation taking into account the photoemission 
final state can clarify the origin of the strong circular dichroism effect of the 
topological surface state.

In conclusion, we observe that the spin polarization measured from the 
in-gap surface state of \BiTe\ and \BiSe\ requires careful cleaving
but that high values not too far from 100\%\ can be obtained
with methods which integrate over macroscopic areas of $\sim0.25$ mm$^2$. 
The degeneracy of surface and bulk states below the Dirac point 
leads, at least for \BiTe, via hybridization to a bulk state which 
has the spin topology of the topological surface state imposed. 
A large circular dichroism effect in the angle distribution of photoelectrons
is measured which, with the help of the spin-resolved data,
 is shown to point along the spin of the electrons. 
The circular dichroism does not require special detectors and 
allows for count rates which are by 2 to 3
orders of magnitude larger than with spin-resolved photoemission. 
It is, therefore, worthwhile investigating by a photoemission calculation
 whether or not the present dichroism 
measures {\it via} the angular momentum indirectly the spin structure. 

\phantom{x}

\noindent



\newpage

\phantom{zeile}

\noindent {\bf Figure captions}

\phantom{zeile}

\noindent Figure 1.  (Color online) Band topology of a spin-orbit split
surface state for (a) the Rashba effect and (b) the topological
insulator for the example of \BiTe.  
(c) Photoemission geometry.
(d) Structure and cleavage plane (not to scale). 
(e)  Brillouin zone and LEED pattern at 80 eV. 
(f) Angle-resolved photoemission at 50 eV photon energy along 
\Gbar-\Mbar. The topological surface state (TSS) appears in the band gap
between the valence-band maximum (VBS) and
the conduction-band mimimum (CBM, taken from 30 eV data) and the two become
degenerate at the Dirac point (\Ed). 
$p$- and $s$-polarized light reveal symmetry properties. 
(g) Constant energy surfaces of the topological surface state 
including the Fermi surface from photoemission at 50 eV photon energy.  

\phantom{zeile}

\noindent Figure 2.   (Color online) 
  Spin- and angle-resolved photoemission of the topological surface state  
for \BiTe\ and \BiSe. 
Spin up ($\bigtriangleup$), spin down  ($\bigtriangledown$),
spin asymmetry (from raw data), and spin polarization 
spectrum (background subtracted) ($\diamond$). 
Note that the spectra in (a) and (b) were taken on opposite sides 
of normal emission (\Gbar). The bulk band gap is marked. 
The magnitude of the spin polarization of the topological surface state feature 
amounts to $(82\pm10)$\%\ for \BiTe\ and $(24\pm15)$\%\ for \BiSe. 
BS, which is predicted to be bulk for \BiTe, is highly spin polarized as well.

\phantom{zeile}

\noindent Figure 3.  (Color online) 
  Circular dichroism of the topological surface state of \BiTe. 
(a) Angle-resolved photoemission intensity $I$ 
for  $\sigma+$ light   and  $\sigma-$  light along with the dichroism  
[$I(\sigma+) - I(\sigma-)$]/[$I(\sigma+) + I(\sigma-)$]. 
(b) Constant-energy cuts at various binding energies $E_{\rm B}$. 
For the topological surface state we see above \Ed\ 
a circular dichroism effect of 20\%\ (at 0.1 eV binding energy).
(c) Dichroism spectra measured for a similar \kpara\ af $E_{\rm F}$ 
[see dashed lines in (a)] as in Fig. 2.
The photon energy is 50 eV. 

\begin{figure}[ht]
	\centering
  \includegraphics[width=1\textwidth]{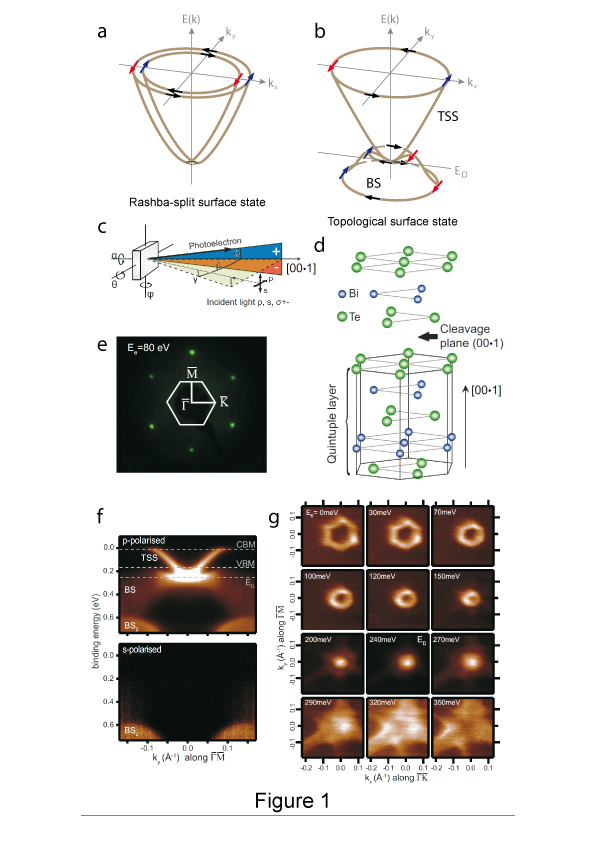}
\end{figure}

\begin{figure}[ht]
	\centering
  \includegraphics[width=1\textwidth]{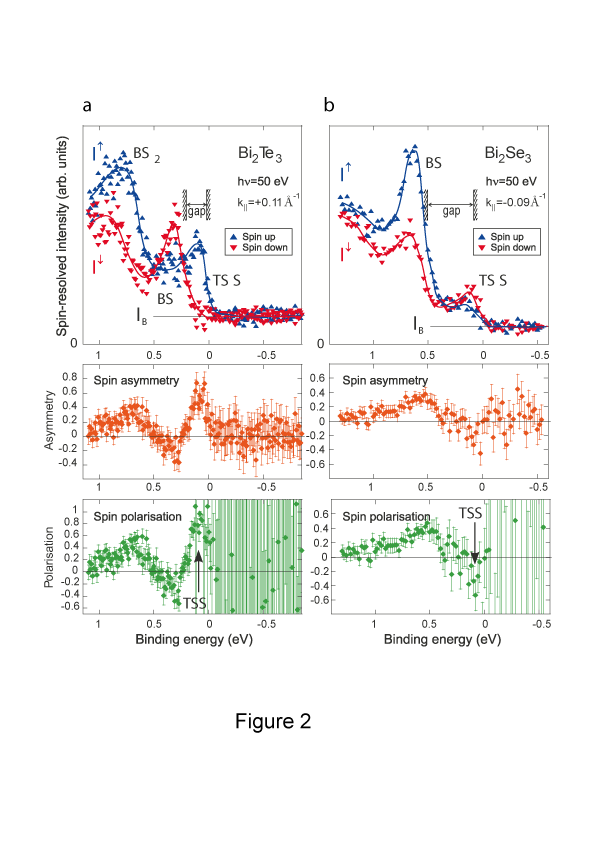}
\end{figure}

\begin{figure}[ht]
	\centering
  \includegraphics[width=1\textwidth]{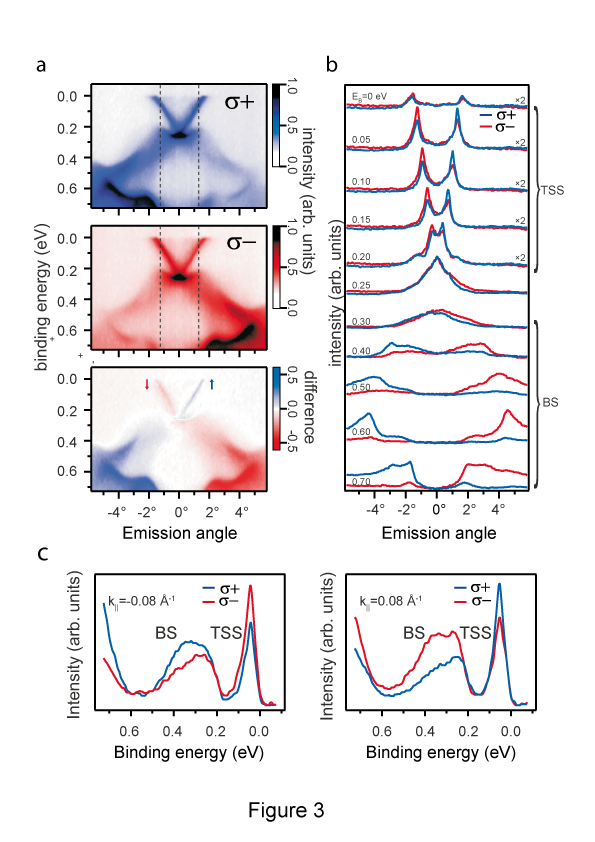}
\end{figure}

\end{document}